\documentstyle[12pt,epsf]{article}
\begin{document}


\newcommand{\nc}{\newcommand}
\nc{\beq}{\begin{equation}}
\nc{\eeq}{\end{equation}}
\nc{\beqa}{\begin{eqnarray}}
\nc{\eeqa}{\end{eqnarray}}
\nc{\lra}{\leftrightarrow}
\nc{\ra}{\rightarrow}
\nc{\sss}{\scriptscriptstyle}
{\nc{\lsim}{\mbox{\raisebox{-.6ex}{~$\stackrel{<}{\sim}$~}}}
{\nc{\gsim}{\mbox{\raisebox{-.6ex}{~$\stackrel{>}{\sim}$~}}}
\def\dsl{\partial\!\!\!/}
\def\lameff{\lambda_{\rm eff}}
\def\Re{{\rm Re\,}}
\def\Im{{\rm Im\,}}
\def\ns{\!\!\!\!\!\!\!\!\!\!\!\!\!\!\!\!\!\!\!\!\!\!\!\!\!\!\!\!\!\!\!\!\!\!\!}
\def\NS{\ns\ns\ns\ns}
\def\dsl{\partial\!\!\!/}
\def\Pl{P_{\sss L}}\def\Pr{P_{\sss R}}
\def\VEV#1{\langle #1 \rangle}
\def\sfrac#1#2{{\textstyle\frac{#1}{#2}}}
\def\ems{{\epsilon_{\rm \bar{MS}} }}
\def\etat{{\eta_t}}
\def\etab{{\eta_b}}
\def\zetat{{\zeta_t}}
\def\zetab{{\zeta_b}}
\def\ttL{{\tilde t_L}}
\def\ttR{{\tilde t_R}}
\def\tbL{{\tilde b_L}}
\def\tbR{{\tilde b_R}}
\def\ttp{{\tilde t_+}}
\def\ttm{{\tilde t_-}}
\def\ttpm{{\tilde t_\pm}}
\def\tbp{{\tilde b_+}}
\def\tbm{{\tilde b_-}}
\def\tbpm{{\tilde b_\pm}}
\def\tqp{{\tilde q_+}}
\def\tqm{{\tilde q_-}}
\def\tqpm{{\tilde q_\pm}}
\def\tqL{{\tilde q_L}}
\def\tqR{{\tilde q_R}}


\begin{titlepage}
\pagestyle{empty}
\baselineskip=21pt
\rightline{McGill/96-20}
\rightline{CERN-TH/96-76}
\rightline{hep-ph/9605235}
\rightline{May 6, 1996}
\vskip .4in

\begin{center}
{\large{\bf Supersymmetric Electroweak Phase Transition:\\
            Beyond Perturbation Theory}}
\end{center}
\vskip .1in

\begin{center}
James M.~Cline

{\it McGill University, Montr\'eal, Qu\'ebec H3A 2T8, Canada,}

and

Kimmo Kainulainen

{\it CERN, CH-1211, Gen\`eve 23, Switzerland.}
\end{center}

\vskip 0.7in

\centerline{ {\bf Abstract} }
\baselineskip=20pt
\vskip 0.5truecm

We compute the three-dimensional effective action for the minimal
supersymmetric standard model, which describes the light modes
of the theory near the finite-temperature electroweak phase transition,
keeping the one-loop corrections from the third generation quarks and
squarks.  Using the lattice results of Kajantie {\it et al.}\ for the
phase transition in the same class of 3-D models, we find that the
strength of the phase transition is sufficient for electroweak
baryogenesis, in much broader regions of parameter space than have been
indicated by purely perturbative analyses.  In particular we find that,
while small values of $\tan\beta$ are favored, positive results persist
even for arbitrarily large values of $\tan\beta$ if the mass of the
$A^0$ boson is between 40 and 120 GeV, a region of parameters which has
not been previously identified as being favorable for electroweak
baryogenesis.
\end{titlepage}
\baselineskip=18pt

\section{Introduction}

One of the fundamental questions in nature is the origin of the
asymmetry of matter over antimatter in the universe.  Although
explanations abound, one of the most interesting possibilities is that
the baryon asymmetry was created during the electroweak phase
transition (EWPT) \cite{review}, using new physics at sufficiently low
energies to be verifiable in anticipated experiments like LEP-II or the
Large Hadron Collider. Although the EWPT is a first order transition in
the standard model, it is too weakly so to fulfill Sakharov's
out-of-thermal-equilibrium requirement for generating baryons:  any
asymmetry created during the EWPT would be quickly erased afterwards by
residual sphaleron interactions in the broken phase of the
$SU(2)_L\times U(1)$ gauge theory \cite{KLRS1}. Moreover it appears
that the standard model has too little CP violation for electroweak
baryogenesis \cite{Gavelaetal}.

It is therefore interesting to find out whether a more strongly first
order EWPT is possible in extensions of the Standard Model, a prime
example being its minimal supersymmetric extension, the MSSM, shown to
be suitable for electroweak baryogenesis in ref.~\cite{HN}.  The
phase transition has been studied in this model by means of the
one-loop finite-temperature effective potential
\cite{GM}-\cite{Delepine}, with the result that there exist some
regions of parameter space where electroweak baryogenesis is possible.
However the perturbative approach should be viewed with skepticism
because at finite temperature it becomes infrared divergent for small
values of the Higgs field, which can be crucial for determining the
critical temperature and Higgs field VEV at the phase transition.

To deal with the breakdown of perturbation theory, Kajantie {\it et
al.} \cite{KLRS1} have studied the EWPT of the Standard Model on the
lattice.  As they have emphasized however, almost any extension of the
Standard Model can  be reduced to an effective three-dimensional theory
of one Higgs doublet interacting with $SU(2)$ gauge bosons, by
integrating out all the modes with thermal masses larger than those of
the longitudinal gauge bosons \cite{KLRS2}.  The beauty of this
approach is that the effective theory need only be numerically studied
once; after that it is simply a matter of matching the parameters of
the fundamental theory onto this effective theory, which can be
reliably done using perturbation theory.

In this paper we construct the effective 3-D Lagrangian corresponding
to the MSSM at finite temperature, keeping the dominant effects
proportional to the top quark Yukawa coupling.  At the critical
temperature it has the simple form
\beq
   \bar{\cal L}_3 = |(\partial_i -
   \sfrac{i}{2}\bar g_3 \vec\tau\cdot\vec W)\Phi|^2 +
   \bar m^2 \Phi^\dagger\Phi +
   \bar \lambda_3 (\Phi^\dagger\Phi )^2 +\frac14 F_{ij}F_{ij}.
\label{e1}
\eeq
where the couplings $\bar \lambda_3$ and $\bar g_3$ depend, ultimately,
on physical parameters of the MSSM such as $\tan\beta$, Higgs boson
masses, and squark masses. The criterion from lattice studies \cite{KLRS1}
for preserving any baryon asymmetry created during the EWPT is
\beq
    {\bar \lambda_3\over \bar g_3^2} < 0.04.
\label{e2}
\eeq

We find that this bound is satisfied in a significant fraction of the
MSSM parameter space, in contrast with previous studies based on a
purely perturbative approach \cite{beqz}.  Although the most recent
perturbative investigations obtained more positive results by
considering negative values of the squark mass parameter $m^2_U$
\cite{carenaetal} or small values of the right-handed squark
\cite{Delepine}, in the present work we find that such choices, while
compatible with electroweak baryogenesis, are not particularly favored
and represent only a small fraction of the total volume of
baryogenesis-allowed parameter space.  The quantities to which our
results turn out to be most sensitive are the ratio of Higgs VEV's,
$\langle H_2 \rangle/ \langle H_1 \rangle = \tan\beta$, the mass of the
pseudoscalar Higgs boson $A^0$ and the soft supersymmetry breaking
squark mixing parameters. We will show that the allowed regions can be
characterized roughly by $0.5 < \tan\beta < 2$ and $m_{A^0}$
unrestricted, or $m_{A^0}$ between 40 and 120 GeV for arbitrarily large
$\tan\beta$, and no special restrictions on the other parameters except
that the largest portion of the allowed space corresponds to large
squark mixing parameters.  This would appear to be a much less
constrained situation than was previously believed to exist for the
MSSM as regards baryogenesis.

The most important interactions affecting the strength of the phase
transition are those involving the largest couplings to the Higgs
field, namely the top quark Yukawa coupling $y_t$. It is conceivable
that $\tan\beta\sim m_t/m_b$, in which case the bottom quark Yukawa
coupling $y_b$ would also be large.  Then the relevant part of the MSSM
Lagrangian, including the neutral sector of the two Higgs fields (so
the $H_i$ below are not doublets) and the third generation quarks and
squarks, can be written in Euclidean space as
\def\leff{{\frac{g^2 + g'^2}{8}}}
\beqa
      {\cal L}_{\rm tree}
       &=& \sum_{i=1,2}\left( |DH_i|^2 + m_i^2|H_i|^2 \right)
        + m_3^2(H_1^*H_2 + {\rm h.c.})
        + \leff (|H_1|^2 - |H_2|^2)^2 \nonumber\\
       &+& y_t\, \bar t_L H_2 t_R + {\rm h.c.\ }
        + y_t^2 |H_2|^2 (|\tilde t_L|^2 + |\tilde t_R|^2 )
        +  y_t\,\tilde t_L^* ( \mu  H_1 + A_t H_2) \tilde t_R
             + {\rm h.c.} \nonumber \\[2 mm]
       &+& y_b\, \bar b_L H_1 b_R + {\rm h.c.\ }
        + y_b^2 |H_1|^2 (|\tilde b_L|^2 + |\tilde b_R|^2 )
        +  y_b\,\tilde b_L^* ( \mu  H_2 + A_b H_1) \tilde b_R
             + {\rm h.c.} \nonumber \\[2 mm]
       &+& \left(\sfrac14 g^2\left(|{\tilde t}_L|^2- |{\tilde b}_L|^2\right)
        - \sfrac{1}{12} g'^2\left(|{\tilde t}_L|^2+|{\tilde b}_L|^2\right)
        +\sfrac{1}{6} g'^2 \left( 2|{\tilde t}_R|^2 -|{\tilde b}_R|^2\right)
        \right) (|H_1|^2 - |H_2|^2) \nonumber\\[2 mm]
       &+& m^2_Q \left(|\tilde t_L|^2+|\tilde b_L|^2\right)
        + m^2_U|\tilde t_R|^2 + m^2_D|\tilde b_R|^2.
\label{e3}
\eeqa
where the $m^2_i$'s for $i=1,2,3$ contain both the soft-breaking and the
supersymmetric contributions.  In the present work we will
consider only moderatly large values for $\tan\beta$, so that
$y_b\ll y_t$ in the numerical analysis; the terms of
order $y_b$ are nevertheless displayed, to allow for future
investigation of the large $\tan\beta$ regime.  However even if $y_b\ll
y_t$, the bottom squarks can still be relevant because of certain loop
diagrams proportional to masses, for which they contribute competetively
with the top squarks if they are sufficiently heavy, and these effects
we do include throughout.

To apply the lattice gauge theory bound (\ref{e2}) one must carry out
two steps \cite{KLRS2}.  First, integrate out all the heavy degrees of
freedom in the finite-temperature theory at the phase transition.  This
means everything except for a single light linear combination of the
Higgs fields, and the transverse gauge bosons, resulting in the
three-dimensional effective action (\ref{e1}) for these fields.
Second, renormalize the same theory at zero temperature so as to
express the parameters appearing in $\bar{\cal L}_3$ as functions of
physical observables, such as particle masses.  In both steps we
compute the corrections due to third generation quarks and squarks
proportional to $y_t$ and $y_b$.  These are diagrams of order $y^2$,
$y^4$ and $g^2 y^2$.  Because the divergences of the theory at zero and
at finite temperature are identical, the 3D Lagrangian parameters are
completely finite and independent of renormalization scale or scheme
when expressed in terms of the physical observables.

\begin{figure}
\hspace{2.0truecm}
\epsfysize=1.5truecm\epsfbox{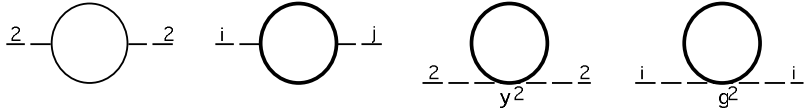}
\vspace{0.1truecm}
\baselineskip 17pt
\caption {The 1PI diagrams needed for the scalar 2-point function.
Thin line in the loop represents quarks and the heavy lines squarks.
Indices labeling external legs refer to doublet; when index is
a letter either 1 or 2 is allowed. The example shown corresponds to
top quark or squark in the loop; for bottom sector reverse 1 and 2.}
\end{figure}
\baselineskip 20pt

\section{Finite-temperature effective lagrangian}

The procedure for reducing (\ref{e3}) to the finite temperature 3-D
theory is straightforward.  One starts with the same 1-loop Feynman
diagrams as at zero temperature; the graphs relevant for the present
problem are shown in figures 1 and 2.  However, at finite $T$, the
integrals over $p_0$ become sums over Matsubara frequencies, $p_0\to
2\pi n T$ for bosons and $p_0\to (2n+1)\pi T$ for fermions, and $\int
d^{4-2\epsilon} p \to 2\pi T\sum_n d^{3-2\epsilon} p$.  The sum goes
over all $n\neq 0$ for the bosons and all $n$ for the fermions to
obtain the effective 3-D theory of the zero Matsubara frequency modes
of the Higgs and gauge bosons, eq.~(\ref{e1}).  Just as for $T=0$, one
can use dimensional regularization (or dimensional reduction in the
case of SUSY) to regulate the ultraviolet divergences; the divergent
counterterms are exactly the same for $T>0$  as for $T=0$.  Defining
$\eta_{t,b}\equiv 3y_{t,b}^2/16\pi^2$,
$L_B = \ln Q^2/(4\pi T)^2 + 2\gamma_E$ and
$L_F = \ln Q^2/(\pi T)^2 + 2\gamma_E$, where $\gamma_E \simeq 0.5772$
and $Q$ is the arbitrary renormalization scale, the resulting finite-$T$
effective Lagrangian is
\beqa
    {\cal L}_3/T
    &=& {\cal L}_{\rm tree}
    \,+\, \etat L_F |\partial H_2|^2
    \,+\, \etab L_F |\partial H_1|^2
     \nonumber\\[1 mm]
    &+& \left(\sfrac{3}{4} y_t^2 T^2
       - \etat L_B (m^2_Q+m^2_U) \right)|H_2|^2
       - \etat L_B |\mu H_1 + A_t H_2|^2
     \nonumber\\ [1 mm]
    &+& \left(\sfrac{3}{4} y_b^2 T^2
       - \etab L_B (m^2_Q+m^2_D) \right)|H_1|^2
       - \etab L_B |\mu H_2 + A_b H_1|^2
     \nonumber\\ [0 mm]
    &+&\frac{g'^2}{32\pi^2}L_B\,
         (m^2_D+m^2_Q- 2m^2_U)(|H_1|^2-|H_2|^2)
     \nonumber \\[1 mm]
    &+&\etat y_t^2 (L_F-L_B)|H_2|^4
     - \sfrac14 \etat (g^2+g'^2) L_B |H_2|^2 (|H_1|^2-|H_2|^2)
     \nonumber\\ [1 mm]
    &+& \etab y_b^2 (L_F-L_B)|H_1|^4
     + \sfrac14 \etab (g^2+g'^2) L_B |H_1|^2 (|H_1|^2-|H_2|^2),
\label{e4}
\eeqa
where we have used the $\overline{\rm MS}$ subtraction scheme, which
means that the combination $1/\epsilon +\ln 4\pi -\gamma_E$ has been
subtracted from divergences.  However the factors of $\ln 4\pi$ and
$\gamma_E$ arise in a different way at finite temperature, which is why
they still appear in the quantities $L_B$ and $L_F$.  Eq.~(\ref{e4}) is
an expansion in $m^2_{Q,U,D}/T^2$ and we have accordingly dropped all
terms of  order $(\mu/T)^2$ and $(A_{t,b}/T)^2$.

\begin{figure}
\vskip -0.3truecm
\hspace{3.5truecm}
\epsfysize=3.5truecm\epsfbox{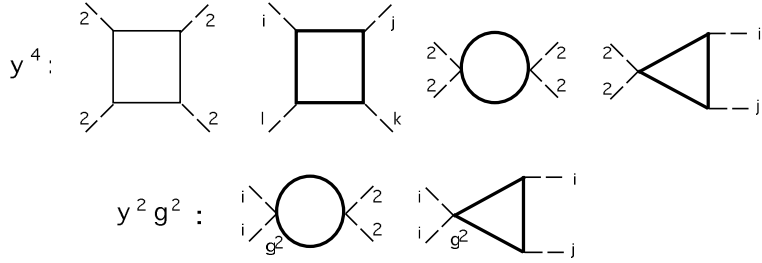}
\vspace{0.3truecm}
\baselineskip 17pt
\caption {The 1PI diagrams needed for the 4-point function.
For explanations see figure 1.}
\end{figure}
\baselineskip 20pt

This is not yet the complete result for $\bar{\cal L}_3$; so
far we have only integrated out the nonzero Matsubara frequency modes,
which have masses of the order $\pi n T$.  However there still remain
particles with masses intermediate between this ``superheavy'' scale,
and the light scale which is of order the magnetic mass of the
transverse gauge bosons ($g^2 T$).  So we have to also integrate out
the zero-Matsubara-frequency modes (called ``heavy'') of the
squarks, gauge bosons, and Higgs bosons, to an accuracy of $y^2$, $y^4$
or $g^2 y^2$ in $\bar{\cal L}_3$. The diagrams that contribute are
identical to the ones we already considered in deriving eq.~(\ref{e4});
the difference is that the heavy particle masses are no longer given by
$2 \pi n T$, but rather the Debye mass.

To integrate out these remaining heavy modes, we must therefore
determine their Debye masses, which consist of a tree-level part plus
a thermal correction.  For the left- and right-handed third-generation
squarks in a vanishing background field ($H_i=0$) one finds \cite{CE}
\beqa
m^2_{\tilde q_L} &=& m_Q^2 + \frac{4g_s^2}{9}T^2
+ \frac{y_t^2+y_b^2}{6}T^2 + \frac{g^2}{4}T^2 +
\frac{g'^2}{108}T^2
\nonumber\\
m^2_\ttR &=& m_U^2 + \frac{4g_s^2}{9}T^2
+ \frac{y_t^2}{3}T^2 + \frac{4g'^2}{27}T^2;
\nonumber\\
m^2_\tbR &=& m_D^2 + \frac{4g_s^2}{9}T^2
+ \frac{y_b^2}{3}T^2 + \frac{g'^2}{27}T^2,
\label{stopmds}
\eeqa
where the contributions from gauginos and charginos have been omitted,
under the implicit assumption that they are so heavy that they
decouple.  (We have checked that including the latter contributions in
the Debye masses has no qualitative effect on our subsequent numerical
results.)  In fact these are the only Debye masses we need because the
squarks are the only particles in ${\cal L}_3$ whose tree-level
couplings are proportional to $y$ or $y^2$. Ignoring the thermal loop
corrections to these couplings, since they would only give two-loop
corrections to $\bar\lambda_3$, the result of integrating out the heavy
modes of the squarks is
\beqa
\bar{\cal L}_3/T &=& {\cal L}_3/T
 + \frac{\zetat}{3 M^2_{D_t}} |\mu \partial_i H_1 + A_t \partial_i H_2|^2
 + \frac{\zetab}{3 M^2_{D_b}} |\mu \partial_i H_2 + A_b \partial_i H_1|^2
\nonumber\\[1.5 mm]
&-& \zetat M^2_{D_t} S^t_+ - \zetab M^2_{D_b} S^b_+ - D_T(|H_1|^2-|H_2|^2)
\nonumber\\ [2.5 mm]
&-& \frac{\zetat y_t^2}{4} \frac{M^2_{D_t}}{m_\ttL m_\ttR} (S^t_-)^2
 -  \frac{\zetab y_b^2}{4} \frac{M^2_{D_b}}{m_\tbL m_\tbR} (S^b_-)^2
\nonumber\\
&-&\frac{\zetat g^2}{8}\frac{M_{D_t}}{m_\ttL}
 \left(1 + \frac{4g'^2}{3g^2}\left( \frac{m_\ttL}{m_\ttR}
  - \frac{1}{4} \right)\right) S^t_-
\left( |H_1|^2 - |H_2|^2 \right)
\nonumber\\
&+&\frac{\zetab g^2}{8}\frac{M_{D_b}}{m_\tbL}
 \left(1 + \frac{2g'^2}{3g^2}\left( \frac{m_\tbL}{m_\tbR}
  - \frac{1}{2} \right)\right) S^b_-
\left( |H_1|^2 - |H_2|^2 \right),
\label{finalL3}
\eeqa
where we defined $M_{D_q} \equiv m_\tqL + m_\tqR$,
$\zeta_q \equiv 3y_q^2 T/4\pi M_{D_q}$,
$S^t_\pm \equiv |H_2|^2 \pm M^{-2}_{D_t}|\mu H_1 + A_t H_2|^2$,
$S^b_\pm \equiv |H_1|^2 \pm M^{-2}_{D_b}|\mu H_2 + A_b H_1|^2$ and
\beq
D_T = \frac{g'^2T}{8\pi}\left(2{m_\ttR}-{m_\tqL} -{m_\tbR}   \right).
\label{DT}
\eeq
The effective lagrangian so obtained is almost in the desired form, but
it still depends on two Higgs doublets rather than one.  At the phase
transition, only one linear combination of the two is massless
($\Phi_l$), while the orthogonal direction is a heavy field ($\Phi_h$)
which must also be integrated out.  But since there are no
self-couplings of the Higgs fields proportional to quark Yukawa
couplings, this final step induces no new terms of the order of $y^4$
or $y^2g^2$ in ${\cal L}_3$; it is just a matter of projecting out the
heavy field.  Let the angle $\alpha$ describe the direction in field
space whose eigenvalue in the temperature-dependent mass matrix of the
two Higgs fields vanishes, at the phase transition temperature $T_c$:
\beq
 \left(\begin{array}{c}  H_1 \\ H_2\end{array}\right)
 = \left(\begin{array}{cc} \cos \alpha & -\sin \alpha
 \\ \sin \alpha & \phantom{-}\cos \alpha \end{array}\right)
 \left(\begin{array}{c} \Phi_l \\ \Phi_h  \end{array}\right).
\label{angle}
\eeq
Then the effect of integrating out the heavy field $\Phi_h$ is simply
to replace $H_1$ by $\cos\alpha\;\Phi_l$ and $H_2$ by $\sin\alpha
\;\Phi_l$.

Further let $m^2_{i,\rm eff}$ be the entries of the Higgs mass matrix,
analogous to the tree-level $m^2_i$'s, but now corrected by the thermal
loop diagrams.  If we define the matrices
\beq
\label{matrixdefs1}
   {\bf m^2} = \left(\begin{array}{cc} m_1^2 & m_3^2\\ m_3^2 & m_2^2\\
   \end{array}\right);\quad
   {\bf A}_t =
           \left(\begin{array}{cc} \mu^2 & \mu A_t\\ \mu A_t & A^2_t\\
   \end{array}\right);\quad
   {\bf A}_b =
           \left(\begin{array}{cc} A^2_b & \mu A_b\\ \mu A_b & \mu^2\\
   \end{array}\right),
\eeq
and
\beq
\label{matrixdefs2}
   {\bf P}_b = \left(\begin{array}{cc} 1 & 0\\ 0 & 0\\
   \end{array}\right);\quad
   {\bf P}_t = \left(\begin{array}{cc} 0 & 0\\ 0 & 1\\
   \end{array}\right);\quad
   {\bf P}_3 = \left(\begin{array}{cc} 1 & 0\\ 0 & -1\\
   \end{array}\right),
\eeq
then ${\bf m^2_{\rm eff}}$ can be written as
\beqa
\label{meff} {\bf m^2_{\rm eff}}
 &=& {\bf m^2} + \Biggl\{
     - \frac12 \etat L_f \left\{ {\bf P}_t,\, {\bf m}^2 \right\}
     - \frac{\zetat}{6M^2_{D_t}} \left\{ {\bf A}_t,\, {\bf m^2}\right\}
     \nonumber\\[1 mm]
 &+& \left( \sfrac34 y^2_t T^2 - \etat L_B (m^2_Q+m^2_U)
       - \zetat M^2_{D_t}\right) {\bf P}_t
   - \left(\etat L_B + \zetat\right){\bf A}_t
     \nonumber\\ [1 mm]
 &+& (t \ra b; \; m^2_U \ra m^2_D) \Biggr\}
    \nonumber\\
 &-&\left( \frac{g'^2}{32\pi^2}L_B\, (2 m^2_U-m^2_D-m^2_Q)
     + D_T\right){\bf P}_3
    \nonumber\\
 &\equiv& {\bf m^2} + \frac{1}{2} \left\{ \delta{\bf Z}_T, {\bf m^2}\right\}
 +{\bf \Pi }_T(0),
\eeqa
where in the last line we made some definitions that will be useful later.
The mixing angle is given by
\beq
   \sin \alpha = \frac{-m^2_{1,\rm eff}}{(m^4_{1,\rm eff}
                                       +  m^4_{3,\rm eff})^{1/2}}
\eeq
at $T_c$, the temperature at which ${\rm Det}(m^2_{\rm eff})=0$.
The anticommutators in eq.~(\ref{meff}) arise due to
wave function renormalization (rescaling $H_i$ so that
the kinetic term in $\bar{\cal L}_3$ is properly normalized).

At one loop the gauge coupling $g_3$ is not renormalized by the Yukawa
couplings. We checked this by computing the correction to $g_3$ from
the correlator $\Phi_l \Phi_l A_iA_i$. The four relevant diagrams are
shown in figure 3.  After rescaling the fields to the canonical
normalization, the direct contributions to this correlator are found to
cancel those induced by wave function renormalization. So even after
the heavy scale integration we have the tree level relation $\bar g^2_3
= g^2T$.

\begin{figure}
\vskip 0.5truecm
\hspace{4truecm}
\epsfysize=1.8truecm \epsfbox{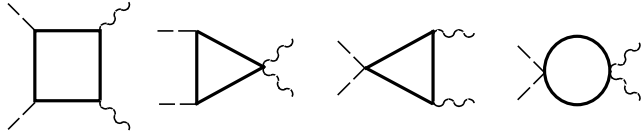}
\vspace{0.3truecm}
\baselineskip 17pt
\caption {Diagrams contributing to the effective gauge coupling
${\bar g}_3$ in the heavy scale (zero-Matsubara-frequency) integration.}
\end{figure}
\baselineskip 20pt

It is now straightforward to extract from eq.~(\ref{finalL3}) the
temperature-corrected quartic coupling of the light doublet $\Phi_l$.
We find that
\beqa
 {\bar\lambda_3\over\bar g^2_3}
 &=& {g^2+g'^2\over 8g^2}\cos^2\!2\alpha
   \nonumber\\
 &+& {3\ln 2\over 4\pi^2}\left({y_t^4\over g^2}\sin^4\!\alpha
        +\frac{g^2+g'^2}{4g^2} y_t^2 \cos 2\alpha\sin^2\!\alpha  \right)
   \nonumber\\
 &+& {3\ln 2\over 4\pi^2}\left({y_b^4\over g^2}\cos^4\!\alpha
        -\frac{g^2+g'^2}{4g^2} y_b^2 \cos 2\alpha\cos^2\!\alpha  \right)
   \nonumber\\
 &-&\frac{3y_t^4}{16\pi g^2}\, \frac{M_{D_t}\, T}
                  {m_\ttL\, m_\ttR}\,(S^t_\alpha )^2
  - \frac{3y_b^4}{16\pi g^2}\, \frac{M_{D_b}\, T}
                  {m_\tbL\, m_\tbR}\,(S^b_\alpha )^2
   \nonumber\\
 &-&\frac{3y_t^2}{32\pi}\,\frac{T}{m_\ttL} \left(1 + \frac{4g'^2}{3g^2}\left(
        \frac{m_\ttL}{m_\ttR} - \frac{1}{4} \right) \right)
     S^t_\alpha \cos 2\alpha
   \nonumber\\
 &+&\frac{3y_b^2}{32\pi}\,\frac{T}{m_\tbL} \left(1 + \frac{2g'^2}{3g^2}\left(
        \frac{m_\tbL}{m_\tbR} + \frac{1}{2} \right) \right)
     S^b_\alpha \cos 2\alpha
   \nonumber\\
 &-& \frac{y_t^2}{16\pi}\,\frac{g^2+g'^2}{g^2}\,\frac{T}{M^3_{D_t}}
      \left(\mu^2 \cos^2\alpha - A^2_t \sin^2\alpha\right)\cos 2\alpha ,
   \nonumber\\
 &+& \frac{y_b^2}{16\pi}\,\frac{g^2+g'^2}{g^2}\,\frac{T}{M^3_{D_b}}
      \left(\mu^2 \sin^2\alpha - A^2_b \cos^2\alpha\right)\cos 2\alpha ,
\label{e5}
\eeqa
where
$S^t_\alpha\equiv \sin^2\!\alpha
- M_{D_t}^{-2}(\mu\cos\alpha + A_t\sin\alpha)^2$ and
$S^b_\alpha\equiv \cos^2\!\alpha
- M_{D_b}^{-2}(\mu\sin\alpha + A_b\cos\alpha)^2$.

Eq.~(\ref{e5}) gives the number that is directly bounded by the lattice
results for the condition that the phase transition be sufficiently
first order, eq.~(\ref{e2}).  However, the renormalization scale
independence of the Lagrangian (\ref{finalL3}), is not yet apparent,
and it contains undetermined parameters which must be expressed in
terms of physical quantities.  Once this is done, ${\cal L}_3$ becomes
manifestly finite and independent of the scale $Q$.

\section{Relation to the physical parameters}

\begin{figure}
\vskip 0.5truecm
\hspace{4.8truecm}
\epsfysize=1.8truecm \epsfbox{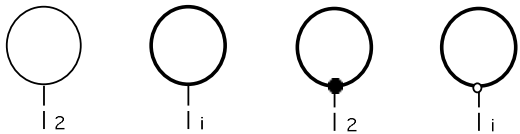}
\vspace{0.3truecm}
\baselineskip 17pt
\caption {The tadpole diagrams needed for the 1-loop effective
potential in the present approximation. The vertices marked by heavy
circles (open circles) come from  quartic terms of order $y^2$ ($g^2$),
where one external Higgs field is replaced by its VEV.}
\end{figure}
\baselineskip 20pt

The next step is to perform the zero-temperature renormalization to the
same accuracy as we did at finite-temperature in order to express the
parameters appearing in eq.~(\ref{e5}) in terms of physical quantities,
namely particle masses and the vacuum expectation values (VEV's) of the
two Higgs fields.  The VEV's are determined by minimizing the 1-loop
effective potential, through the equations
\beqa
&& \hskip - 0.9truecm
m^2_1 + \tan \tilde\beta \; m^2_3 + \frac{g^2+g'^2}{4}
(\tilde v^2_1-\tilde v^2_2) \; + \;
\frac{1}{2 \tilde v_1}\frac{{\rm d}V_{1-\rm loop}}{{\rm d}\tilde h_1} = 0,
\nonumber \\ && \hskip - 0.9truecm
m^2_2 + \cot \tilde\beta \; m^2_3 - \frac{g^2+g'^2}{4}
(\tilde v^2_1-\tilde v^2_2) \; + \;
\frac{1}{2\tilde v_2}\frac{{\rm d}V_{1-\rm loop}}{{\rm d}\tilde h_2} = 0,
\label{mincond}
\eeqa
where we have split the Higgs fields into CP-even and odd parts, $H_i
= \tilde h_i + i \tilde\chi_i$, and $\tilde v_i = \langle
H_i \rangle= \langle \tilde h_i \rangle$.  The ratio of VEV's is
$\tan\tilde\beta = \tilde v_2/\tilde v_1$.  Because we have not yet
accounted for wave function renormalization  at one loop, the
$\tilde v_i$'s are not the physical VEV's, defined to be $v_i =
\langle  h_i \rangle$, but the two sets of fields are related by matrix
equations
\beq
\tilde h = ({\bf Z}_h)^{1/2} h \cong (1 + \sfrac12 \delta{\bf Z}_h) h;
\qquad \tilde \chi = ({\bf Z}_\chi)^{1/2} \chi \cong
(1 + \sfrac12 \delta{\bf Z}_\chi) \chi.
\eeq
The matrices $\delta{\bf Z}$ can also be expressed in terms of the
derivative of the 1-loop vacuum polarizations of the fields,
$\delta{\bf Z} = ({\rm d}{\bf\Pi}(p^2)/{\rm d}p^2)|_{p^2=0} =
{\bf\Pi}'(0)$.

\begin{figure}
\vskip -0.3truecm
\hspace{2.5truecm}
\epsfysize=3.5truecm\epsfbox{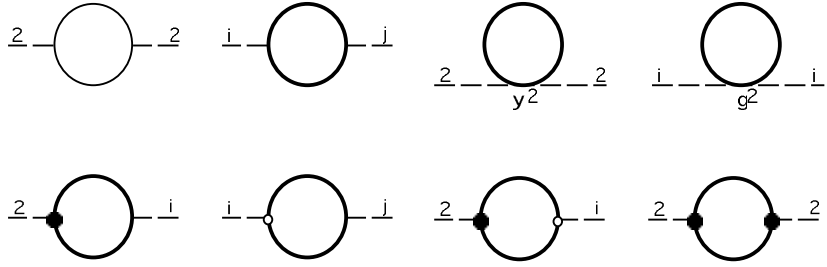}
\vspace{0.3truecm}
\baselineskip 17pt
\caption
{Diagrams contributing to the irreducible two point functions of the
higgs fields in the broken phase. Vertices induced by spontaneous
symmetry breaking do not contribute to the CP-odd two-point function.
Again the indices on the external legs correspond to the top sector.}
\end{figure}
\baselineskip 20pt

The minimization conditions (\ref{mincond}) give us two equations for
the three parameters $m^2_i$.  To determine the third we must compute
the physical mass of one of the Higgs bosons.  A convenient choice is
the CP-odd scalar, $A^0$.  Its pole mass, $m_A$, is determined by
\beq
{\rm Det} ( \sfrac12 {\bf V}_{\!\!,\chi\chi} +
{\bf \Pi}_{\chi}(m^2_{A})-m^2_{A} ) = 0,
\label{detequ1}
\eeq
where $ \sfrac12{\bf V}_{\!\!,\chi\chi} =
{\partial}^2V_0/{\partial}\chi_i{\partial}\chi_j$
is the tree-level mass matrix,
\beq
 \sfrac12 {\bf V}_{\!\!,\chi\chi} =
    {\bf m}^2 + \frac{g^2 + g'^2}{4}(\tilde v^2_1 - \tilde v^2_2)
    \left(\begin{array}{cc} 1 & 0  \\0 & -1 \end{array}\right).
\label{massmatrix}
\eeq
Of course an analogous expression could be used to relate the parameters
to the CP-even Higgs masses, but the CP-odd sector is simpler because
it contains a massless particle, the Goldstone boson.  Rather than solving
for the exact pole mass, we will renormalize at $p^2=0$, which
means expanding ${\bf \Pi}_{\chi}(m^2_{A}) \simeq
{\bf \Pi}_{\chi}(0) + {\bf \Pi}'_{\chi}(0)m^2_{A}$, so that
${\bf \Pi}_{\chi}(m^2_{A})+m^2_{A}$ becomes ${\bf\Pi}_{\chi}(0)
- m^2_A{\bf Z}_{\chi}^{-1}$.
After some manipulations using  eq.~(\ref{mincond}) to eliminate the
second term of (\ref{massmatrix}), and using the explicit form of
$dV_{1-\rm loop}/{\rm d}h_i$ (see the appendix), eq.~(\ref{detequ1})
can be written as
\beq
{\rm Det} \left( \begin{array}{cc}
       \Delta \tan \beta_\chi - m^2_{A} & -\Delta   \\
      -\Delta  & \Delta \cot \beta_\chi - m^2_{A}
          \end{array} \right) = 0,
\label{detequ2}
\eeq
where
\beq
\Delta \equiv - {m^2_{\chi 3}} + \etat \mu A_t F_t(Q) +
\etab \mu A_b F_b(Q),
\label{delta}
\eeq
$ m^2_{\chi 3}$ is the off-diagonal element of the symmetric matrix
${\bf  m}^2_\chi \equiv {\bf m}^2
+\sfrac12 \{\delta {\bf Z}_\chi, {\bf m}^2\}$,
$F_{t,b}(Q)$ are the corrections due to squark loops,
\beq
 F_q(Q) =  \sum_\pm {\pm m^2_\tqpm \over
        m^2_\tqp - m^2_\tqm } \left(\ln \frac{Q^2}{m^2_\tqpm }
        +1\right),
\label{ahva}
\eeq
and the squark masses are given by \cite{Ellisetal}
\beqa
m^2_\ttpm &=& \sfrac12(m^2_U + m^2_Q) + m^2_t + \sfrac14 M^2_Z\cos 2\beta
\nonumber\\
&& \pm \sqrt{\sfrac14[m^2_Q-m^2_U + \sfrac12 C_tM^2_Z \cos 2\beta]^2
+m^2_t (A_t+\mu \cot \beta )^2};\nonumber
\eeqa
\beqa
m^2_\tbpm &=& \sfrac12(m^2_D + m^2_Q) + m^2_b - \sfrac14 M^2_Z\cos 2\beta
\nonumber \\
&& \pm \sqrt{\sfrac14[m^2_Q-m^2_D - \sfrac12 C_b M^2_Z\cos 2\beta]^2
+ m^2_b (A_b + \mu \tan \beta )^2},
\label{stopmass}
\eeqa
where $C_t = 1-\sfrac83 \sin^2\theta_{\rm W}$
and $C_b = 1-\sfrac43 \sin^2\theta_{\rm W}$.

To achieve the simple form (\ref{detequ2}) for the $A^0$ pole mass
condition, we had to introduce the shifted angle $\beta_\chi$, defined by
$\tan\beta_\chi \equiv v_{\chi 2}/v_{\chi 1}$, where
$v_{\chi i} = (Z^{-1/2}_\chi)_{ij} \tilde v_j$.  The relation to the
physical $\tan\beta$ is scale-independent:
$
\tan\beta_\chi = \tan\beta \left (1
     -\sfrac12\Delta Z_{11}  + \sfrac12\Delta Z_{22}
     + \Delta Z_{12} \cot 2\beta \right),
$
where $\Delta{\bf Z} = {\bf Z}_h - {\bf Z}_\chi$.  In the $\overline
{\rm MS}$ subtraction scheme, which we are using throughout,  the wave
function renormalization matrices of the CP-odd and CP-even fields
$\chi$ and $h$ are
\beqa
\delta {\bf Z}_\chi &=&
                -\etat \; \ln \frac{Q^2}{m^2_t} \;{\bf P}_t
                -\etat H(m^2_\ttp,m^2_\ttm) {\bf A}_t
\; + \; (t\ra b); \nonumber\\
\label{wfr}
\delta {\bf Z}_h &=& \delta {\bf Z}_\chi
\nonumber\\
&+& \etat\left(\frac{2}{3} - \frac{m^2_t(m^2_\ttp + m^2_\ttm)}
{3m^2_\ttp m^2_\ttm} \right){\bf P}_t
+ \frac{\etat}{6}m_t\sin 2\theta_t
     \frac{m^2_\ttp-m^2_\ttm}{m^2_\ttp m^2_\ttm} {\bf P}_{A_t}
\nonumber\\
&+& \etat\sin^22\theta_t\left(H(m^2_\ttp,m^2_\ttm )
-\frac{m^2_\ttp+m^2_\ttm}{12m^2_\ttp m^2_\ttm} \right){\bf A}_t
\nonumber\\
&+& \frac{\etat}{24}M^2_Z\sin 2\beta
 \left( \frac{m^2_\ttp + m^2_\ttm}{m^2_\ttp m^2_\ttm }
+ C_t \cos 2\theta_t \frac{m^2_\ttp - m^2_\ttm}{m^2_\ttp m^2_\ttm }
 \right){\bf B}_t\nonumber\\
&+& \frac{3m_t}{128\pi^2}(g^2+g'^2) \sin 2\theta_t
\left(\frac{m^2_\ttp - m^2_\ttm}{12 m^2_\ttp m^2_\ttm } - C_t\cos 2\theta_t
\left( H(m^2_\ttp,m^2_\ttm ) -
\frac{m^2_\ttp + m^2_\ttm}{12 m^2_\ttp m^2_\ttm }
\right)\right) {\bf C}_t\nonumber\\
&+& (t\ra b),
\label{dZCpeven}
\eeqa
with ${\bf P}_{1,2}$ and ${\bf A}_{t,b}$ defined in (\ref{matrixdefs1}
-\ref{matrixdefs2}) and
\beqa
{\bf P}_{A_t} = \left( \begin{array}{cc} 0 & \mu  \\ \mu & 2A_t
     \end{array}\right)\;;&&
{\bf P}_{A_b} \;=\;\left( \begin{array}{cc} 2A_b & \mu  \\ \mu & 0
     \end{array}\right); \nonumber\\
{\bf B}_t = \left( \begin{array}{cc} 0 & -1  \\ -1 & 2\tan \beta
     \end{array}\right)\;;&&
{\bf B}_b \;=\; \left( \begin{array}{cc} 2\cot \beta & -1  \\ -1 & 0
     \end{array}\right)\nonumber\\
{\bf C}_t = \left( \begin{array}{cc} 2 \mu\cot\beta & A_t\cot\beta-\mu
 \\  A_t\cot\beta-\mu & -2A_t
     \end{array}\right)\;;&&
{\bf C}_b = \left( \begin{array}{cc} - 2A_b & A_b\tan\beta-\mu
 \\ A_b\tan\beta-\mu & 2\mu \tan \beta
     \end{array}\right).
\label{matrixdefs3}
\eeqa
The squark mixing angles are defined by
$\sin 2\theta_t = 2 m_t (A_t + \mu \cot \beta )/(m^2_\ttp -m^2_\ttm )$
and $\cos 2\theta_t = (m^2_U - m^2_Q - \sfrac12 C_t m^2_Z \cos 2\beta)/
(m^2_\ttp -m^2_\ttm )$ (for the bottom squarks let $t\to b$, $m^2_U\to
m^2_D$ and $\cos\beta\leftrightarrow\sin\beta)$
and
\beq
 H(m^2_+,m^2_-) \equiv \frac{1}{2(m^2_+ - m^2_-)}\left(
                   \frac{m^2_+ + m^2_-}{m^2_+ - m^2_-}
                 - \frac{2m^2_+ m^2_-}{(m^2_+ - m^2_-)^2}
                   \ln \frac{m^2_+}{m^2_-}\right).
\label{phi}
\eeq

Equation (\ref{detequ2}) has a vanishing eigenvalue corresponding to the
Goldstone boson, and the nonzero eigenvalue is the mass of the $A^0$,
\beq
m^2_A = \frac{2\Delta}{\sin 2\beta_\chi}.
\label{mphys}
\eeq
Our procedure for computing $m^2_A$ is somewhat more complicated than
that of ref.~\cite{Ellisetal} which parametrized all the effects
of wave function renormalization in scale-dependent VEV's, $v_i(Q^2)$,
whereas we take the VEV's and hence $\tan\beta$ to be physical,
scale-independent quantities.  Our answer reduces to theirs if we
neglect the wave function renormalization effects, i.e.\ by taking
$\beta_\chi\to\beta$ and $m^2_{\chi 3}\to m^2_3$.

The solution (\ref{mphys}) allows the lagrangian mass parameter $m^2_3$
to be expressed in terms of $m^2_A$. Then, having found $m^2_3$, the
other elements $m_{1,2}^2$ are determined by the minimization
conditions (\ref{mincond}). It is convenient however, to give the
results in terms of the scaled quantity $m^2_{\chi 3}$ 
(see below eq.~(\ref{delta})). We find:
\beqa
{\bf m}^2_\chi (Q) &=& \sfrac12 m^2_A \sin 2\beta_\chi
\left( \begin{array}{cc} \tan \beta_\chi  & -1   \\
      -1  & \cot \beta_\chi \end{array} \right)
\nonumber \\[1 mm]
&+& \etat F_t(Q) {\bf A}_t + \etab F_b(Q) {\bf A}_b
 +  \etat G_t(Q) {\bf P}_t  +\etat G_b(Q) {\bf P}_b
\nonumber\\[1 mm]
&-&\sfrac12 M^2_Z ( Z^h_{11} \cos^2\beta - Z^h_{22} \sin^2\beta )
                ({\bf P}_b Z^\chi_{11} - {\bf P}_t Z^\chi_{22})
\nonumber\\
&+& \sfrac{3}{128\pi^2} (g^2+g'^2) D(Q) {\bf P}_3,
\label{m3m1}
\eeqa
where
\beq
 G_q(Q) \equiv \sum_\pm m^2_\tqpm \left(\ln {Q^2\over
        m^2_\tqpm }+1\right) - 2m^2_q \left(
        \ln \frac{Q^2}{m^2_q} + 1 \right)
\label{geku}
\eeq
and the other new function, coming from the tadpoles of the
$g^2|\tilde q|^2 |H|^2$ terms of eq.~(\ref{e3}), is given by
\beqa
D(Q) &=&  \sum_\pm m^2_{\tilde t\sss\pm}
\left(\ln (Q^2/m^2_{\tilde t\sss\pm})+1\right) -
\sum_\pm m^2_{\tilde b\sss\pm}
\left(\ln (Q^2/m^2_{\tilde b\sss\pm})+1\right) \nonumber\\
&&+ C_t(m^2_Q-m^2_U + \sfrac12 C_t M^2_Z \cos 2\beta) F_{t}(Q)
\nonumber\\
&&- C_b(m^2_Q-m^2_D - \sfrac12 C_b M^2_Z \cos 2\beta) F_{b}(Q).
\label{deku}
\eeqa

These are all the expressions needed to solve for the entries of the
effective mass matrix ${\bf m}^2_{\rm eff}$. It is convenient to do so
in terms of the scaled matrix ${\bf m}_\chi$, using eq.~(\ref{e4}):
\beq
{\bf m}^2_{\rm eff} = {\bf m}^2_\chi
+ \frac{1}{2} \{ \delta {\bf Z}_T - \delta {\bf Z}_\chi,
{\bf m}^2_\chi \} + {\bf \Pi}_T(0),
\label{effm}
\eeq
where $\delta {\bf Z}_T$ and ${\bf \Pi}_T(0)$ were defined in
(\ref{meff}).
The difference between the wave function renormalization matrices at
finite and at zero temperature is finite and scale-independent, as
it must be. Moreover, the $Q$-dependence appearing in ${\bf m}_\chi^2$
(\ref{m3m1}) is precisely what is needed  to cancel that coming from
${\bf \Pi}_T(0)$ at the accuracy to which we are working,
so that the matrix ${\bf m}^2_{{\rm eff}}$ and hence the angle
$\cos^2\!\alpha$ are scale-independent.  The result is
\beqa
{\bf m}^2_{\rm eff}
&=& {\bf m}^2_\chi (T)
 + \Biggl\{ (2 \etat c_B -\zetat ){\bf A}_t
 + \left(\sfrac34 y_t^2T^2 + 2\etat c_B (m^2_Q+m^2_U)
        -\zetat M^2_{D_t}\right) {\bf P}_t
  \nonumber\\ [1 mm]
&+& \frac12 \left( \etat H(m^2_\ttp ,m^2_\ttm)
          - \frac{\zetat}{3M^2_{D_t}}\right)
          \left\{ {\bf A}_t, {\bf m^2} \right\}
 +  \frac12 \etat \left( 2 c_F + \ln\frac{T^2}{m^2_t} \right)
          \left\{ {\bf P}_t, {\bf m^2} \right\}
  \nonumber\\ [2 mm]
&+& (t \ra b;\; m^2_U \ra m^2_D) \Biggr\}
+ \left( \frac{g'^2}{16\pi^2} c_B\, (2 m^2_U-m^2_D-m^2_Q)
     - D_T\right){\bf P}_3,
\label{finalmasses}
\eeqa
where the matrix ${\bf m}^2_\chi (T)$ is as in (\ref{m3m1}) but with
$Q=T$, and we define $c_B \equiv
\ln 4\pi - \gamma_E$ and $c_F \equiv \ln \pi - \gamma_E$.  To one-loop
accuracy, it suffices to use the tree level expression for ${\bf m}^2$
appearing in the anticommutators, which in terms of physical parameters
is given by
\beq
{\bf m}^2_{\rm tree} = \sfrac12 m^2_A \sin 2\beta
\left( \begin{array}{cc} \tan \beta  & -1   \\
      -1  & \cot \beta \end{array} \right)
- \sfrac12 M^2_Z \cos 2\beta\; {\bf P}_3.
\label{mtree}
\eeq

In the next section we will identify the regions of parameter space
where baryogenesis is allowed.  One must check whether these parameters
are in fact compatible with other constraints, including the
experimental lower limit on the mass of the lightest Higgs boson $h^0$,
whose tree level expression vanishes when $\tan \beta = 1$.  Since we
will find that low values of $\tan \beta$ are relevant for electroweak
baryogenesis, it is important to include loop corrections to $m_{h^0}$,
which can be large.  The pole mass of the $h^0$ is determined similarly
to eq.~(\ref{detequ1}):
\beq
{\rm Det} (\sfrac12 {\bf V}_{\!\!,hh} - {\bf \Pi}_h(m^2_h)-m^2_h ) = 0,
\label{detequ3}
\eeq
where
\beq
 \sfrac12 {\bf V}_{\!\!,\chi\chi} =
    {\bf m}^2 + \frac{g^2 + g'^2}{4}
    \left(\begin{array}{cc} 3\tilde v^2_1 - \tilde v^2_2 &
                           -2\tilde v_1\tilde v_2  \\
                           -2\tilde v_1\tilde v_2 &
                            3\tilde v^2_2 - \tilde v^2_1
\end{array}\right).
\label{massmatrix2}
\eeq
Expanding the 2-point function $\Pi_h(m^2_h)$ and performing other
manipulations similar to the CP-odd case, eq.~(\ref{detequ3}) can be
put into the form
\beq
{\rm Det} ({\bf M}^2(\beta) - m^2_h ) = 0,
\label{detequ4}
\eeq
where
\begin{eqnarray}
{\bf M^2}(\beta ) &=& \frac{1}{2} m^2_A \sin 2\beta_\chi
\left( \begin{array}{cc} \tan \beta_\chi  & -1   \\
       -1  &  \cot \beta_\chi \end{array} \right)
+ \sfrac12 M^2_Z\sin 2\beta
\left( \begin{array}{cc} \cot \beta  & -1   \\
       -1  &  \tan \beta \end{array} \right)
\nonumber\\
&+&
\frac{\etat}{2}\sin^22\theta_t \; I(m^2_\ttp/m^2_\ttm) \;
{\bf A}_t + 2\etat m^2_t \ln \frac{m^2_\ttp m^2_\ttm}{m^4_t} {\bf P}_t
\nonumber \\
&+& \etat m_t \sin 2\theta_t \ln \frac{m^2_\ttp}{m^2_\ttm} {\bf P}_{A_t}
+ (\sfrac12 M^2_Z \sin 2\beta\; \delta Z^h_{22} + \Pi^t_1(Q))
{\bf B}_t \nonumber\\
&+& \Pi^t_2 \; {\bf C}_t  \; \; + \; \; (t \ra b),
\label{monster}
\end{eqnarray}
with
\beq
I(r) = 2 - \frac{r+1}{r-1}\ln r
\label{gr}
\eeq
and
\beqa
\Pi^q_1(Q) &=& \frac{1}{2} \eta_q M^2_Z\sin 2\beta\left(
\ln \frac{Q^2}{m_\tqp m_\tqm}
+ \frac{1}{2} C_q \cos 2\theta_q \ln \frac{m^2_\tqp}{m^2_\tqm}
\right);
\nonumber\\
\Pi^q_2 &=& \frac{3m_t}{256\pi^2}(g^2+g'^2)\; \sin 2\theta_t
\left(\ln\frac{m^2_\ttp}{m^2_\ttm} - C_t\cos 2\theta_t
I(m^2_\ttp/m^2_\ttm) \right).
\label{Pig2y2}
\eeqa
It is easy to see that the $Q$-dependence in (\ref{monster}) due to
$\Pi^t_1(Q)$ ($\Pi^b_1(Q)$) exactly cancels that of $\delta Z^h_{22}$
($\delta Z^h_{22}$).  We verified that the solution of
eq.~(\ref{detequ4}), giving $m^2_h$ as a function of $\tan \beta$,
agrees with previous results \cite{LN}.

\section{Results}

Having completely determined the quantity $\bar{\lambda}_3/\bar{g}^2_3$
in terms of the physical parameters, we now turn to the question of
whether the baryons created during the electroweak phase transition can
be preserved from subsequent sphaleron interactions within the MSSM.
Specifically, for what values of the physical parameters does
$\bar{\lambda}_3/\bar{g}^2_3$ satisfy the bound (\ref{e2})?  {\it A
priori} there seems to be no simple, analytic answer to this question.
No single term among the many loop corrections in eq.~(\ref{meff})
could be identified as being dominant over the others, and furthermore
the parameter space is large: $\tan\beta$, $m_{A^0}$, $m^2_Q$, $m^2_U$,
$m^2_D$, $\mu$, $A_t$, $A_b$.  We therefore chose to do a Monte Carlo 
search of this space for values which satisfy the constraint (\ref{e2}).

The preceding list of parameters does not include the gauge or Yukawa
coupling constants because these can be defined through the tree level
relations, $(g^2+g'^2)/g^2 = m^2_Z/m^2_W$, $y^2_t =
m^2_t/v^2\sin^2\!\beta$ and $y^2_b = m^2_b/v^2\cos^2\!\beta$.  For the
gauge couplings this is a consequence of the vanishing of the Yukawa
contributions to the beta function at one loop, and our neglect of the
difference between the pole masses and the masses defined at zero
external momentum.  As for the Yukawa couplings, these appear only in
loop corrections for us, so to one-loop accuracy it is consistent to
use their tree-level values.

\begin{figure}
\vskip -0.3truecm
\hspace{1.5truecm}
\epsfysize=18truecm\epsfbox{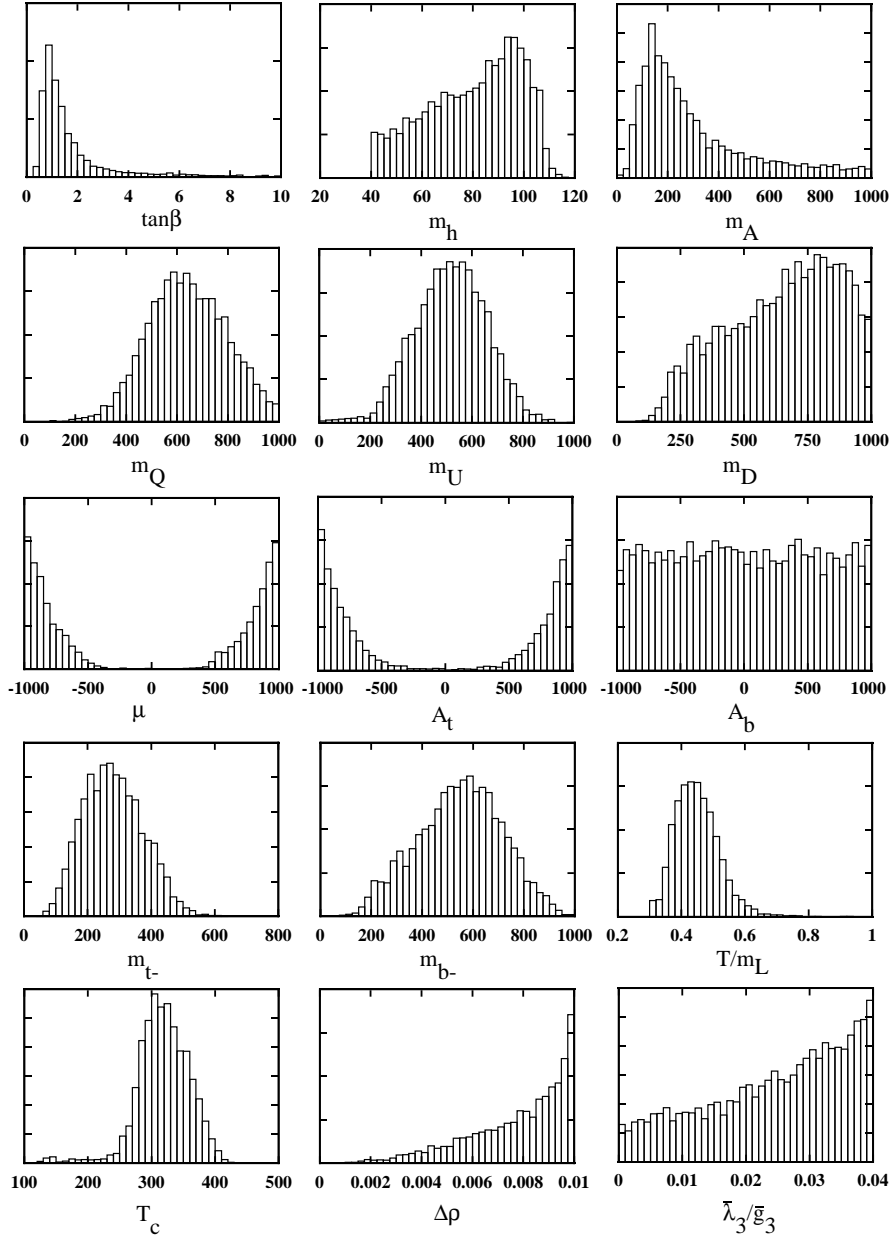}
\vspace{0.3truecm}
\baselineskip 17pt
\caption {The projected distributions of the data satisfying the sphaleron
washout bound (\ref{e2}) obtained from the Monte Carlo run described in
the text.  Units are GeV.}
\end{figure}
\baselineskip 20pt

For the Monte Carlo search we found it convenient to take as
independent parameters those listed in Table I, which shows the ranges
over which they were varied.  The massive parameters were allowed to be
as large as 1 TeV.  These MC-generated sets were subjected to various
constraints.  The lower limits for the pseudoscalar mass $m_{A^0}$ and
the squark masses were taken to be 20 GeV and 45 GeV, respectively, and
we used the recent top quark mass measurement of $m_t=175$ GeV
\cite{topmass}.  We further required that the squark masses and mixings
be consistent with deviations of the $\rho$ parameter from unity of
less than 0.01 (see ref.~\cite{beqz}).  Specifically, the contributions
from the top-bottom quark and squark splittings as computed in
ref.~\cite{Lim}, including squark mixing effects, were constrained to
satisfy $\Delta\rho(t,b)+\Delta\rho(\tilde t,\tilde b)<0.01$.  Finally,
the accepted data were required to satisfy the baryogenesis constraint
(\ref{e2}) and stability of the potential ($\bar \lambda_3 > 0$).  The
distributions of the parameters within this set, as well as histograms
of derived quantities like the squark and Higgs masses,
$\bar{\lambda}_3/\bar{g}^2_3$, and the critical temperature, are shown
in figure 6.  As a rough indication of how special the 5,600 accepted
sets were among all possibilities, including some with unphysical
masses or couplings, we needed 40 million trials to generate them.  The
baryogenesis-allowed cases thus constitute approximately 0.014 percent
of the full parameter space.

\def\phm{\phantom{-}}
\begin{table}
\begin{center}
\begin{tabular}{|c|c|c|c|c|c|c|}
\hline \hline
$\tan\beta$ & $m_{A^0}$ & $m^2_{Q,U,D}$     & $\mu, A_t, A_b$  \\ \hline
    0.4     &  20 GeV   &   0               &    $-$1 TeV      \\
    10      &  1 TeV    &   1 TeV$^2$       &  \phm 1 TeV      \\
\hline\hline
\end{tabular}
\end{center}
\noindent Table 1: Minimum and maximum values used in the Monte Carlo
of the parameters.
\end{table}

One of the most striking features of the distributions is that
$\tan\beta$ is sharply peaked near unity and falls to a small but
constant value of approximately $10^{-2}$ of the maximum frequency.
Thus it would appear that small values of $\tan\beta$ are strongly
favored by our results.  This is somewhat misleading however, because
there is a strong correlation between $\tan\beta$ and $m_{A^0}$, as
shown in figure 7.  The probability of getting large values of
$\tan\beta$ is very much dependent upon $m_{A^0}$.  If for whatever
reason it became known that $m_{A^0}$ was in the region of $40-120$ GeV,
the distribution of $\tan\beta$ would be much flatter than is shown in
figure 6, with very large values being almost as likely as small ones.
This correlation is also evident in the distribution for $m_{A^0}$,
which jumps up at small values, due to the enlargement of allowed
parameters in the direction of increasing $\tan\beta$.  

Moreover the distributions for $\mu$ and $A_t$ show a very clear
preference for large mixings.  There are allowed parameters also close
to $\mu, A_t =0$, but those corresponding to large mixing are much more
numerous.  The large squark mixing angles are also correlated with the
rather large average value of 300 GeV of the critical temperature.  Due
to the smallness of the coupling $y_b$, the bottom squark sector
corrections are small, which shows in the flatness of the
$A_b$-distribution.

\begin{figure}
\vskip -0.5truecm
\hspace{-0.3 truecm}
\epsfysize13truecm\epsfbox{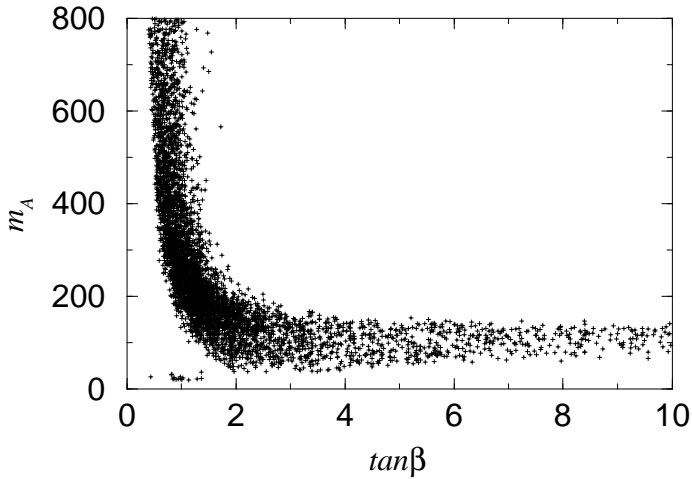}
\vspace{-6truecm}
\baselineskip 17pt
\caption { The baryogenesis-allowed points in the $\tan\beta$-$m_{A^0}$
plane. The sphaleron bound (\ref{e2}) pushes the solutions close to axes; 
the region away from the axes is populated by points with 
$\bar\lambda_3/\bar g^2_3 \gg 0.04$, in violation of the bound. Units
of $m_{A^0}$ are GeV.} 
\end{figure}
\baselineskip 20pt

The other variables also display certain preferences.  The lighter
squark masses peak at 260 GeV and 480 GeV, showing some preference for
a moderately light top squark, though not as light as that advocated by
the perturbative study in reference \cite{Delepine}.
Furthermore the probability for $m^2_U$ to be negative, suggested in
ref.~\cite{carenaetal} as an optimum choice for strengthening the phase
transition, appears to be quite small.
The mass of the lightest Higgs boson, $m_{h^0}$, is in the range of 40
GeV (the experimental lower limit we imposed) to 130 GeV.  The
$\Delta\rho$-distribution \cite{Lim} sharply increases for large
$\Delta \rho$, showing the severity of this constraint on the data.  It
is important to note, however, that large squark mixing angles make it
much easier to satisfy the $\rho$-parameter constraint.  Finally, the
parameter characterizing the strength of the phase transition,
$\bar{\lambda}_3/\bar{g}^2_3$, is monotonically increasing, reflecting
the difficulty of getting a strongly first order transition.

\begin{figure}
\vskip -1.5truecm
\hspace{1.0truecm}
\epsfysize=11.0truecm\epsfbox{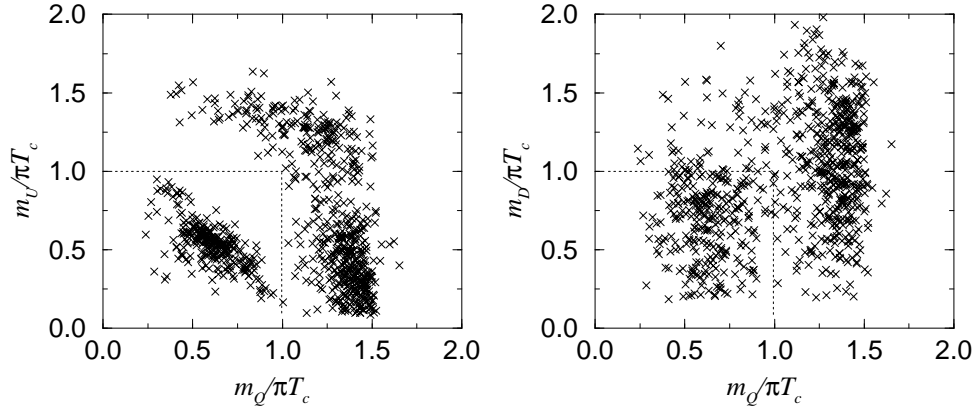}
\vspace{-4truecm}
\baselineskip 17pt
\caption {The density distributions in ($m_Q/\pi T_c,m_U/\pi T_c$) and 
($m_Q/\pi T_c,m_U/\pi T_c$) planes.  Dotted lines show the borders of 
the accepted regions $m_{Q,U,D}/\pi T_c < 1$.}
\end{figure}
\baselineskip 20pt

There are two further constraints that we imposed on our accepted data
sets.   First, care must be taken to ensure that the finite-temperature
perturbative expansion is not breaking down where we need it. In
particular, the thermal squark loop contributions as written in
eq.~(\ref{e4}) are only correct if the zero-Matsubara-frequency
(``heavy'') modes are much lighter than the nonzero ones
(``superheavy''), which requires that $m_{Q,U,D} < 2 \pi T$.  We
checked that the squark contribution to the quartic piece of ${\cal
L}_3$ in eq.\ (\ref{e4}) differs from the exact result (not expanded in
$m^2/T^2$) by less than 5\% for $m_{Q,U,D} < \pi T$ and by a factor of
$\sim 2$ at $m_{Q,U,D} \sim 2 \pi T$.  We imposed the more
stringent cut of $m_{Q,U,D} < \pi T_c$ on our data, which reduced the
size of the sample roughly by a factor of six.  The effect of these
cuts is clearly seen in figure 8, where we show the density plots of
the distributions $m_{Q,U,D}/\pi T_c$, based on a separate run with the
constraint $m_{Q,U,D}<2\pi T_c$. 

Second, one has to insure the validity of the heavy scale perturbation
theory. A typical expansion parameter for integrating out the heavy
modes is $\zeta_t = 3y^2_t T/4\pi M_{D_t}$, which should not become too
large; we required that $T/m < 1$ for all Debye masses m.  From the
distribution for $T/m_{\tilde t_L}$ in figure 6 one sees however that
the other cuts (in particular the requirement $m_{Q,U,D}< \pi T_c$)
already confine the heavy scale expansion parameters to the range
needed to insure $\zeta<1$. This is partly because we did not consider
negative values for $m^2_U$ in the present work.

Clearly, making these consistency cuts can lead us to neglect parameter
values that might actually be acceptable for electroweak baryogenesis.
However it is difficult to consistently implement the dimensional
reduction program except for the relatively light squark masses that
pass our cuts, or in the opposite limit of $m_{Q,U,D} \gg 2 \pi T_c$
where the heavy modes decouple, since only then is there a clear
hierarchy between the superheavy and heavy scales. (Of course the same
problem exists in the purely perturbative effective potential
approach.)  A naive way to interpolate between these limits would be to
replace the $m^2/T^2$ expansions with the exact expressions for the
corresponding finite temperature integrals.  While not quite rigorous,
this might provide a reasonable approximation to the exact result.  An
investigation along these lines is in progress.

\section{Conclusions}

We examined the strength of the first order electroweak phase
transition in the MSSM with respect to the prospects for safeguarding
electroweak baryogenesis from washout by residual sphaleron
interactions, finding rather encouraging results.  Although our
calculations were perturbative, the method---computing the
three-dimensional finite-temperature effective action of the light
Higgs and gauge fields at the phase transition---enabled us to take
advantage of nonperturbative results that have been obtained from
lattice gauge theory computations.  This is therefore the first study
of the phase transition in the MSSM that can claim to be free from the
infrared divergences that make the usual perturbative calculations
untrustworthy.  Indeed, we find that the results of these two distinct
approaches disagree.

The most serious constraint on electroweak baryogenesis in the MSSM,
seen also in the earlier investigations \cite{beqz}-\cite{Delepine},
seems to be the bias\footnote{This restriction is somewhat alleviated
in a two-loop computation, however \cite{Esp}.} toward small values of
$\tan\beta\lsim 2$.  But in our approach this {\it bias} is very
different from a prohibition on large values of $\tan\beta$; in fact
such values are not ruled out, but they must appear in conjunction with
low values ($40-120$ GeV) of the $A^0$ boson mass, as is clear from our
figure (7).  Since there is no intrinsically ``correct'' integration
measure for the space of all parameters in the MSSM, the large $\tan
\beta$ possibility can hardly be considered less natural, even if it
comprised a smaller subset of our Monte Carlo results. If the limit on
$m_{A^0}$ should be improved such as to exclude this region, then it
will become a question of whether $\tan\beta\lsim 1.5$ is viable.

There is one important caveat to our conclusions: as emphasized in
\cite{KLRS2}, the approach used here assumes that the only light
degrees of freedom at the phase transition are the transverse gauge
bosons and a single linear combination of the two Higgs fields.  It is
possible that other fields which are generically heavy happen to also
be light for special parameter values--for example, the Debye masses of
the squarks can vanish if $m^2_{Q,U,D} < 0$.  In such cases we can say
nothing until the lattice computations are redone to take into account
these potential new sources of infrared divergences.

\section* {Note Added}
Soon after this paper was completed,
we received two articles where the same problem was considered.  Losada
\cite{Losada} derived the dimensionally reduced lagrangian, keeping also
the $g^4$-corrections (but setting $g'= 0$).  Laine \cite{Laine} made a
complete analysis in an approximation similar to ours.  Where
comparison is possible, our results are in good agreement.

\section* {Acknowledgements}
We wish to thank Peter Arnold, Gian Giudice,  Keijo Kajantie, Mikko Laine,
Mariano Quiros and Misha Shaposhnikov for useful discussions.

\appendix
\section{Appendix}
Here we give the derivatives of
the effective potential and the zero-momentum limits of 2-point
functions, which were necessary to derive, but not to finally express,
the relevant quantities in the body of the text.
\beqa
\frac{1}{2v_1}\frac{{\rm d}V_1}{{\rm d}h_1}
&=&
-\etat (\mu^2 + \mu A_t \tan \beta )F_t(Q)
-\etab (A^2_b + \mu A_b \tan \beta )F_b(Q)
\nonumber\\
&& - \etab G_b(Q) - \sfrac{3}{128\pi^2} (g^2+g'^2) D(Q)
\nonumber\\
\frac{1}{2v_2}\frac{{\rm d}V_1}{{\rm d}h_2}
&=&
- \etat (A_t^2 + \mu A_t \cot \beta )F_t(Q)
- \etat (\mu^2 + \mu A_b \cot \beta )F_b(Q)
\nonumber\\
&& - \etat G_t(Q) + \sfrac{3}{128\pi^2} (g^2+g'^2) D(Q),
\label{effpot}
\eeqa
with $F_q(Q)$ defined in (\ref{ahva}), $G_q(Q)$ in (\ref{geku}) and
$D(Q)$ in (\ref{deku}).
The two-point function of the CP-odd sector at the zero momentum is
given by
\beqa
{\bf \Pi}(0)^\chi &=& - \etat G_t(Q){\bf P}_t - \etat F_t(Q) {\bf A}_t
\; + \; (t \ra b) \nonumber\\
&&- \sfrac{3}{128\pi^2} (g^2+g'^2) D(Q) {\bf P}_3,
\label{1PICPodd}
\eeqa
with ${\bf P}_i$ defined in (\ref{matrixdefs2}).
In the CP-even sector we find the result
\beqa
{\bf \Pi}(0)^h &=&  {\bf \Pi}(0)^\chi
\nonumber\\
&+& \frac{\etat}{2}\sin^22\theta_t \; I(m^2_\ttp/m^2_\ttm ) \;{\bf A}_t
 + \eta_t m_t \sin 2\theta_t \; \ln \frac{m^2_\ttp}{m^2_\ttm} {\bf P}_{A_t}
\nonumber\\[-1.5 mm]
&+& 2\eta_t \; m^2_t \ln\frac{m^2_\ttp m^2_\ttm}{m^4_t} {\bf P}_t
+ \Pi^t_1(Q) {\bf B}_t +  \Pi^t_2 \; {\bf C}_t
\nonumber\\[1 mm]
&+& (t \ra b),
\label{1PICPeven}
\eeqa
with ${\bf P}_{Aq}$, ${\bf B}_i$ and ${\bf C}_i$ defined in
(\ref{matrixdefs3}), $I(r)$ in (\ref{gr}) and $\Pi^q_i$
in (\ref{Pig2y2}).

\end{document}